\documentclass[10pt,journal,comsoc]{IEEEtran}
\usepackage{url}
\usepackage{caption}
\usepackage[noadjust]{cite}
\usepackage{graphicx}
\usepackage{multirow}
\usepackage{caption}

\usepackage{subcaption}
\usepackage{comment}
\usepackage{xspace}
\usepackage{caption}
\usepackage{subcaption}
\usepackage{makecell}
\usepackage{ragged2e}
\usepackage{rotating}
\usepackage{lscape}
\usepackage[table]{xcolor} 
\usepackage{nomencl}
\usepackage{mwe}
\makenomenclature
\usepackage{wrapfig}
\usepackage{makecell}
\usepackage{balance}
\usepackage[utf8]{inputenc}
\usepackage{ragged2e}
\usepackage{booktabs}

\usepackage{pifont}
\usepackage{tabulary}
\usepackage{comment}
\usepackage{lipsum}
\usepackage{array, tabularx, multirow, booktabs, diagbox}
\usepackage{adjustbox}

\usepackage{amsmath,amssymb,amsfonts}
\usepackage{algorithmic}
\usepackage{graphicx}
\usepackage{textcomp}
\def\BibTeX{{\rm B\kern-.05em{\sc i\kern-.025em b}\kern-.08em
    T\kern-.1667em\lower.7ex\hbox{E}\kern-.125emX}}

\begin{document}
\title{Metaverse for Wireless Systems: Vision, Enablers, Architecture, and Future Directions}

\author{Latif~U.~Khan,~Zhu~Han,~\IEEEmembership{Fellow,~IEEE},~Dusit~Niyato,~\IEEEmembership{Fellow,~IEEE},~Mohsen~Guizani,~\IEEEmembership{Fellow,~IEEE},~and~Choong~Seon~Hong,~\IEEEmembership{Senior~Member,~IEEE}
        
    
\IEEEcompsocitemizethanks{
\IEEEcompsocthanksitem Latif~U.~Khan, and M~Guizani are with the Machine Learning Department, Mohamed Bin Zayed University of Artificial Intelligence, United Arab Emirates.
\IEEEcompsocthanksitem C.~S.~Hong is with the Department of Computer Science \& Engineering, Kyung Hee University, Yongin-si 17104, South Korea.
\IEEEcompsocthanksitem Zhu Han is with the Electrical and Computer Engineering Department, University of Houston, Houston, TX 77004 USA.
\IEEEcompsocthanksitem D. Niyato is with the School of Computer Science and Engineering, Nanyang Technological University, Singapore.

}}

\markboth{}{}%

\maketitle






\begin{abstract} 
Recently, significant research efforts have been initiated to enable the next-generation, namely, the sixth-generation ($6$G) wireless systems. In this article, we present a vision of the metaverse toward effectively enabling the development of $6$G wireless systems. A metaverse uses virtual representation (e.g., digital twin), digital avatars, and interactive experience technologies (e.g., extended reality) to assist analyses, optimizations, and operations of various wireless applications. Specifically, the metaverse can offer virtual wireless system operations through the digital twin that allows network designers, mobile developers, and telecommunications engineers to monitor, observe, analyze, and simulate their solutions collaboratively and virtually. We first introduce a general architecture of metaverse for wireless systems. We discuss key driving applications, design trends, and key enablers of the metaverse for wireless systems. Finally, we present several open challenges and their potential solutions.

\end{abstract}



\section{Introduction}
\begin{figure*}[!t]
	\centering
	\captionsetup{justification=centering}
	\includegraphics[width=18cm, height=8cm]{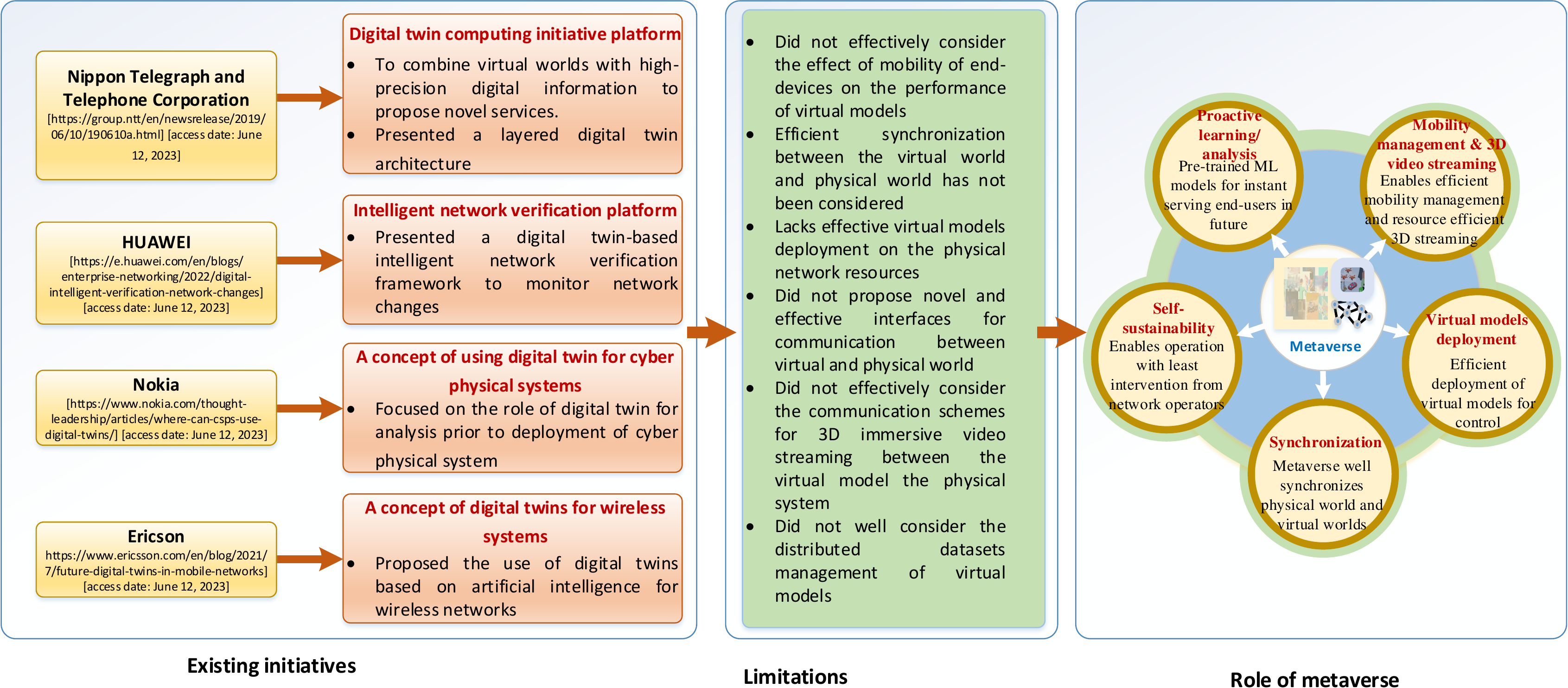}
	\caption{Digital twin use cases, limitations, and the role of metaverse in wireless systems.}
	\label{fig:twinusecases}
\end{figure*}

\setlength{\parindent}{0.7cm}Emerging next-generation wireless system (e.g., the sixth-generation ($6$G) wireless system) applications, such as brain-computer interaction, smart tourism, and industry $4.0$, will be based on diverse requirements and user-defined characteristics \cite{8869705,khan20206g}. To meet such demands, $6$G must adopt and possess special new features such as self-configuring and proactive learning/intelligence \cite{khan2022digital}. A self-configuring wireless system refers to an efficient operation with minimum possible intervention from end-users/network operators. Proactive learning is necessary to optimally utilize network resources (e.g., computing, communication, and energy resources) in response to highly dynamic environments and stringent application requirements. One promising solution to address the design challenges of $6$G systems is a digital twin~\cite{9711524}. The digital twin uses a virtual model of the physical system together with classical approaches such as mathematical optimization, stochastic analysis, and machine learning for performance and cost optimization. Although a digital twin can offer a number of benefits and is effectively applicable to many applications, its use for wireless systems can face many issues especially to meet a variety of mobile service and application requirements. For instance, on one hand, a digital twin can be used to proactively and preemptively analyze a wireless system by using its virtual model. However, without incorporating the effect of mobile users or devices controlled by humans, e.g., unmanned aerial vehicles (UAVs), within the virtual twin model, we may not be able to obtain results (e.g., estimating wireless channels or caching decisions at the network edge) that are accurate for actual physical systems. An overview of existing initiatives using digital twins, their limitations, and the role of metaverse for wireless applications is shown in Fig.~\ref{fig:twinusecases}. In one way, software-defined networking (SDN) technology can be used for the efficient management of network functions. However, SDN cannot directly offer proactive, intelligent analytics and self-sustainability, and thus might not be able to meet the diverse requirements of emerging wireless applications and mobile users.\par

In a metaverse-based system, one can use the concept of digital avatars that model human behavior. Likewise, other mobile devices (e.g., autonomous vehicles (AVs) and UAVs serving and/or loosely supervised by end-users) should also be represented by avatars (i.e., digital copies) in that they are related directly to humans. On the other hand, static entities and objects (e.g., roads, traffic signs, and buildings) should be included in the digital twin model, e.g., they can affect wireless propagation and user mobility characteristics. Therefore, we will need to combine digital twins with avatars and interactive experience technologies in the metaverse to effectively model the physical world of wireless systems. For instance, consider accident reporting of cars and AVs. Roadside units (RSUs) and other objects in the environment can be represented by digital twins. The metaverse will capture the chaos occurred if a real accident happens, e.g., how people, AVs, and other cars react to the situation, in which abnormal and hotspot wireless traffic will be generated and mobility patterns will be observed. Accordingly, interactive experience technologies can be used by metaverse to superimpose additional information on twin models and avatars in meta space. The information can be used by machine learning modules and/or human network administrators to analyze/control physical entities (e.g., UAV flying base stations and edge servers) for reporting accidents and other functions (e.g., lane change assistance and collision avoidance).\par    

The metaverse will offer virtual worlds that facilitate interactions between virtual models (i.e., digital twins) and avatars (i.e., human models) in which the key technologies are augmented reality (AR), virtual reality (VR), extended reality (XR), and mixed reality (MR)~\cite{ning2021survey}. Additionally, the metaverse will use emerging technologies including blockchain, machine learning, network slicing, semantic web, computer vision, and natural language processing to analyze/operate physical systems. In the literature, a few works consider metaverse and its applications~\cite{ning2021survey, huynh2022artificial, ynag2022fusing}. In \cite{ning2021survey}, the authors presented the concept of a metaverse, recent advances, potential applications, and open challenges. The work also highlights social aspects and hyper spatiotemporal views that are important dimensions of the metaverse. The authors in \cite{huynh2022artificial} reviewed the role of artificial intelligence in the metaverse. They outlined six various technical aspects (i.e., neural interface, digital twin, networking, blockchain, machine vision, and natural language processing) towards the development of the metaverse. The authors in \cite{ynag2022fusing} surveyed the fusion of blockchain and metaverse as well the recent advances. In contrast to the works in \cite{ning2021survey,huynh2022artificial,ynag2022fusing}, this article focuses on an application of metaverse to wireless systems, i.e., a metaverse-based wireless system. Note that there are two design aspects: (a) wireless for metaverse and (b) metaverse for wireless. The wireless for metaverse deals with the wireless resource management for efficient metaverse signaling over a wireless network \cite{chang20226g,tang2022roadmap}. On the other hand, the metaverse for wireless deals with the use of metaverse for efficient resource management to enable various applications. Here, our focus is on the metaverse for wireless in contrast to \cite{chang20226g,tang2022roadmap}. We discuss the key enablers as well as key requirements of such a system. Furthermore, we present its architecture along with a wireless system example. The contributions of this paper can be summarized as follows.

\begin{itemize}
    \item We present an overview and our vision of the metaverse towards enabling wireless systems. We are the first to identify the adoption of the emerging metaverse technology, and its potential to improve wireless systems and mobile services through the key enablers, such as interactive experience technologies, avatars, and digital twins.
    \item Following our proposed concept of metaverse for wireless systems, we devise and introduce a general architecture to support its development. We highlight the main requirements that necessitate components and their connections presented in the  architecture. To demonstrate the applicability and usability, we discuss an example scenario of the metaverse for wireless systems.
    \item Finally, we present several future directions that require careful and holistic investigation, analysis, and designs to realize the proposed systems to achieve the full benefits.
\end{itemize}\par

\section{Concept and Key Enablers}
\label{Concept and Key Enablers}
\subsection{Metaverse and Wireless Systems}
\label{Concept}
A metaverse of wireless system will combine digital twins with interactive experience technologies (e.g., AR/VR/MR/XR) and digital avatars to replicate and actuate a physical wireless system \cite{9711524,tao2018digital,khan2022digital}. The first step in the creation of a metaverse is to virtually model (i.e., digital twinning) the physical scenario. The modeling can be performed by using various techniques, including mathematical modeling, simulation modeling, experimental modeling, and data-driven modeling. Next, additional information is superimposed (e.g., virtual objects in mobile environments) on the virtual model by using various sensing and monitoring techniques. Finally, digital avatars of humans are created in the metaverse to model the metaverse-based wireless systems which can represent a variety of stakeholders from users and network operators\footnote{In the rest of the paper, we  use the term ``avatar" to refer to a digital avatar}. A metaverse for a wireless system has two main aspects: offline analysis and online control. The former can guide us regarding the design and deployment of wireless systems. Such an analysis can be performed by using various simulation tools a priori (e.g., for system-level simulation, signal processing, network protocols and routing, and mobility simulations). For a metaverse-based wireless system, new simulators may need to be designed and developed that can  analyze wireless systems more effectively in meta space. We discuss more details about architecture (i.e., physical space and meta space) in Section~\emph{Metaverse-based wireless system architecture and use cases}. In addition to analyzing the virtual wireless world using metaverse, one can use metaverse for run-time control of physical systems. Moreover, to allow wireless metaverse accessible anywhere anytime, wireless communications and networking can be adopted in addition to other technologies (e.g., blockchain and edge computing) for enabling seamless interaction between the virtual world (i.e., meta space) and the physical system (i.e., physical space). Such interactions occur through wireless interfaces. In Section~\emph{Metaverse-based wireless system architecture and use cases}, we discuss how to enable metaverse over a wireless network. Note that a metaverse is different from a digital twin in that,  unlike a digital twin which only uses a virtual model, a metaverse uses a virtual model, avatars\footnote{https://beebom.com/metaverse-avatars-explained/}, and interactive experience technologies for analyzing/ controlling the wireless system.  \par 


\subsection{Enablers}
\label{Enablers}
The key enablers of a metaverse are digital twins, digital avatars, and interactive experience technologies (i.e., AR/VR/MR/XR), and  their roles and wireless counterparts are given in Table~\ref{tab:enablers}. These key enablers are in turn activated by various artificial intelligence (AI) schemes, computing, communication, digital modeling, sensing, and localization technologies. For a wireless metaverse, the first step is to create a digital twin of the physical wireless system. The virtual model of the physical system can be designed using various techniques (e.g., mathematical modeling and experimental modeling) \cite{khan2022digital}. For mathematical modeling, generally, simplified assumptions (e.g., linear approximation for non-linear functions) are made. One can also use experimental modeling that is carried out by a series of experiments. However, there are some scenarios (e.g., wireless propagation in unconventional communications scenarios) that are inefficient or infeasible to be accurately modeled by mathematical modeling and even by experimental modeling \cite{elayan2021digital}. To address this issue, data-driven modeling, which is based on training a machine learning model by using data generated, can be used. Such a machine learning model is based on centralized training or distributed training. Centralized training can offer fast convergence but at the cost of loss of privacy due to the migration of device data to a centralized cloud for training. To address this issue, distributed training, such as federated learning (FL), can be used \cite{khan2021dispersed2}. Distributed training is based on training local models at devices and then sending the local models to the global aggregation server to obtain the global model. However, distributed training-based machine learning has a few challenges, such as non-independent and identically distributed (non-IID) data as well as device and network heterogeneity (i.e., the variable computing capacity of distributed training nodes). Other than virtual modeling for twins, avatars and interactive experience technologies are required for the metaverse. Examples of applications of MR are industrial plant maintenance and healthcare systems. XR covers all three interactive experience technologies, such as AR, VR, and MR, with the aim of making the digital world indistinguishable from the actual world. Additionally, XR provides us with seamless interaction among AR, VR, and MR. The key challenges of XR are interoperability and the design of interfaces for interaction among various interactive experience technologies. Although interactive experience technologies use a virtual representation of humans as avatars in many scenarios (e.g., video games), in the context of a wireless system, novel digital avatars need to be designed that can produce realistic effects (e.g., due to wireless signal energy absorption, wireless signal reflection/refraction, and wireless signal attenuation) as actual humans. For instance, consider Terahertz (THz) communication which generally requires line of sight (LoS) communication. LoS communication in THz communication will be affected by humans in real-time systems \cite{han2019terahertz}. As another example, the concentration of red blood cells (RBCs) affects THz communication-based nanonetworks \cite{salem2018effect}. The path-loss and molecular noise decrease with an increase in the concentration of RBCs, and vice versa. A few other applications based on human-computer interaction (e.g., 3D printing) also require an effective model of avatars. To model avatars, 3D modeling schemes can be used that will capture real-world human characteristics. To enable a wireless system using metaverse using its key enablers, next, we present a general architecture.     \par

\begin{table}

\caption {Metaverse: Enablers, their key role and wireless system counterparts.} \label{tab:enablers} 
  \centering
  \begin{tabular}{p{1.7cm}p{2cm}p{4.3cm}}
    \toprule 

     \textbf{Enabler} & \textbf{Key role} & \textbf{Example in metaverse-based wireless system}   \\
     \midrule
     Interactive experience technologies & To annotate the virtual models for wireless system applications. & AR-based annotations for industry maintenance systems, VR/AR-based entertainment systems, XR-based research testbeds (e.g., Illinois Extended Reality testbed).   \\
     \midrule
     Digital avatars & To enable effects (e.g., mobility, signal loss) of humans of the actual world in metaverse. & Users in autonomous cars model, mobile user models, and unmanned aerial vehicle models controlled by humans.   \\
     \midrule
     Digital twins &  To virtually model the physical wireless system.  &  Base stations and their environment virtual model, smart factory model, and healthcare system model.  \\

\bottomrule 
\end{tabular}
\end{table}

\begin{figure*}[!t]
	\centering
	\captionsetup{justification=centering}
	\includegraphics[width=18cm, height=11cm]{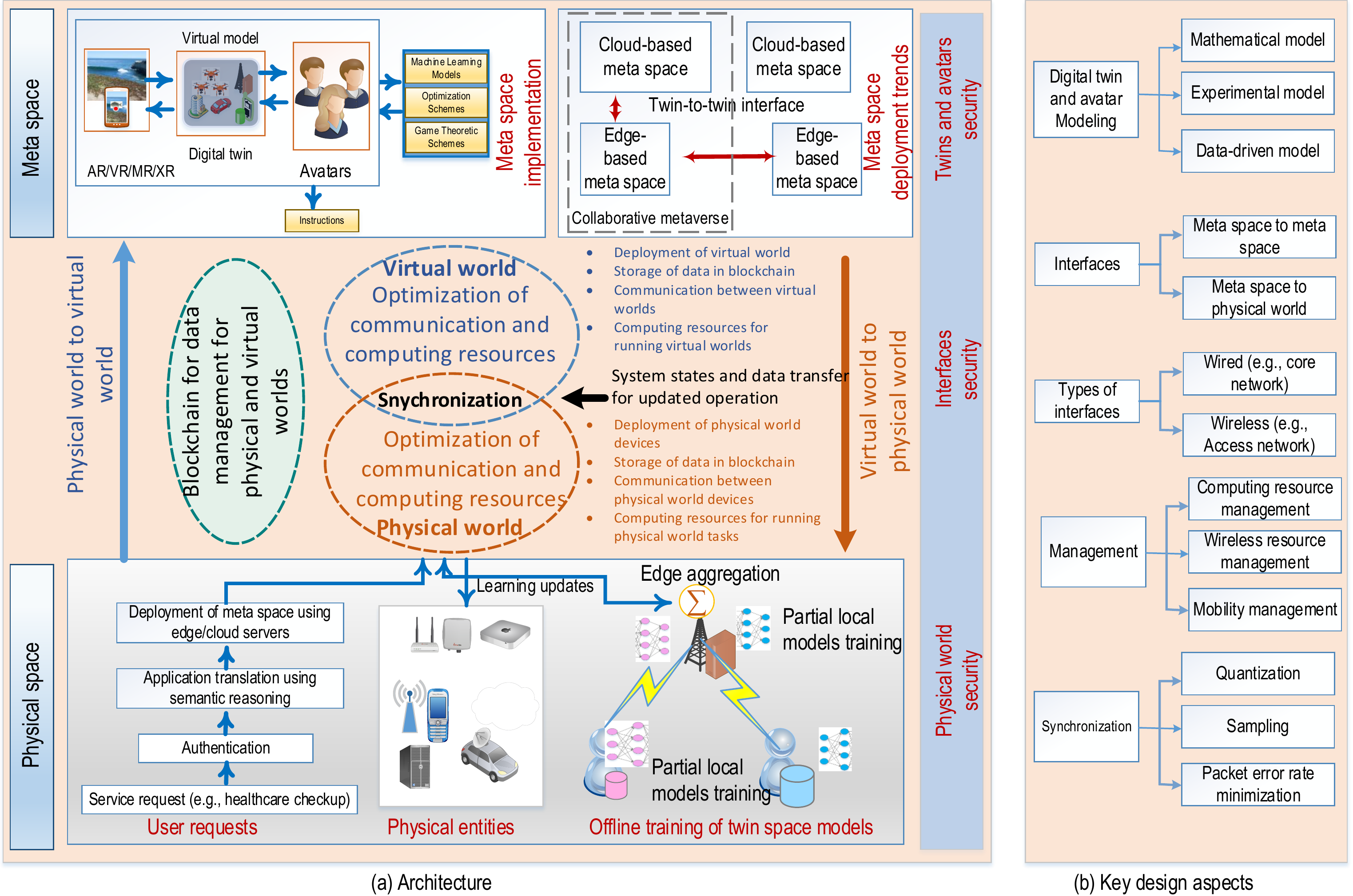}
	\caption{A high-level architecture of a metaverse-based wireless system.}
	\label{fig:architecture}
\end{figure*}

\section{Metaverse-Based Wireless System Architecture and Use Cases}
\label{High-Level Architecture}
\subsection{Architecture}

\subsubsection{Implementation Aspects}
There are three main implementation aspects of the metaverse-based wireless systems, namely, implementations of the meta space, physical interaction space, and interfaces for the communication between the meta space and physical interaction space, as shown in Fig.~\ref{fig:architecture}. A meta space can be implemented using an edge or a remote cloud, depending on the application. For instance, autonomous car functions (e.g., accident reporting) may require instant computing power in a meta space. Therefore, meta space will be implemented using edge servers. On the other hand, the cloud can be used for the implementation of meta space for applications that are delay tolerant. Implementation of a meta space will require seamless interaction among digital twins, avatars, and interactive experience technologies. For the interaction between the meta space and physical space, wired and wireless interfaces can be used. Wired interfaces can use the backhaul links that connect edge devices with the meta space implemented at the remote cloud. On the other hand, the end-devices can be connected with the edge-based metaverse using a radio access network.  Efficient multi-access schemes with effective resource allocation will be required for radio access networks for multiple metaverse-based applications. Wireless resources  will be used for performing various tasks, such as the transfer of data, learning model updates, interactive experience data, and control information. Also, metaverse-based applications must be isolated from each other for seamless, cost-efficient operation over the shared physical network resources (e.g., edge servers and wireless resources). 
\subsubsection{Operation Aspects}
For a metaverse-based wireless system, there are three main operation aspects, such as training of meta space models prior to requests, online control of physical space entities, and end-user requests. Offline training for getting pre-trained meta space models prior to user request can enable proactive learning for efficient resource optimization. To do so, one can use various learning schemes: centralized ML and FL. For a metaverse-based system, FL can be preferably used due to its inherent feature of better privacy preserving capability and low communication resources requirement for taking into account the frequently generated data. For instance, autonomous cars generate 4,000 gigaoctet of data everyday, therefore, transferring the autonomous cars data to the centralized location for training a centralized ML model will consume significant communication resources. To address this issue, one can use FL that only sends local models (i.e., that have significantly lower size than the whole local dataset) to the edge/cloud server for aggregation as shown in Fig.~\ref{fig:architecture}. After getting a pre-trained model using FL, one can store the model using blockchain which will yield immutability and transparency. Such pre-trained models will be used by a meta space to serve end-users upon request. Note that blockchain consensus algorithms generally have computational complexity and energy consumption. Some of the blockchain consensus algorithms (e.g., proof of work) have high latency and energy consumption, whereas few algorithms have  low energy (e.g.,
delegated proof of stake) and low latency (e.g., Byzantine fault tolerance). Therefore, one must design novel algorithms for the metaverse to enable a good tradeoff between latency and energy consumption. Additionally, blockchain has privacy concerns as well due to its distributed nature. Blockchain nodes have access to the data and thus might cause privacy leakage. Therefore, there is a need for privacy-aware blockchain consensus algorithms. \par
Now, we discuss how to make a user request from a metaverse-based wireless system. A user can initiate a request that must be authenticated prior to processing it for avoiding malicious users from accessing the system. Next to authentication, one must translate the user request using semantic reasoning schemes to enable a general meta space for various applications. Using semantic reasoning schemes for the translation of user requests from different domains (e.g., intelligent transportation system and healthcare) to a homogenized form will enable to design of a general meta space for various applications, and thus minimizes the development time and complexity. Next to translation, there is a need to deploy meta space. One can use edge, cloud or both edge and cloud to implement meta space, as shown in Fig.~\ref{fig:architecture}. An edge-based meta space will offer low latency and thus be preferable for use in strict latency applications (e.g., healthcare). Although edge-based meta space can offer low latency compared to cloud-based meta space, it will suffer from the limitation of low storage and computing capacities. To address this, we can deploy meta space using both edge and cloud (i.e., hybrid meta space). After the on-demand deployment of meta space, there is a need to use pre-trained meta space models, mathematical optimization, game theory, and graph theory, among others, to control physical space entities for effectively serving the end-users. Note that on-demand meta space can be implemented using virtual machines and containers. This fashion of creating on-demand virtual machines/containers will enable us to run meta space for the requested service and then release the space after the service time. There may be multiple requests from various applications. To service multiple users of different applications, there is a need to deploy meta space for every application. However, deploying virtual machines/containers for various applications at edge/cloud requires careful design and resource management. Therefore, there is a need to propose novel resource management schemes while deploying meta spaces. Next, we discuss various key requirements for implementing the metaverse to enable wireless systems. \par
\begin{figure}[!t]
	\centering
	\captionsetup{justification=centering}
	\includegraphics[width=8cm, height=6cm]{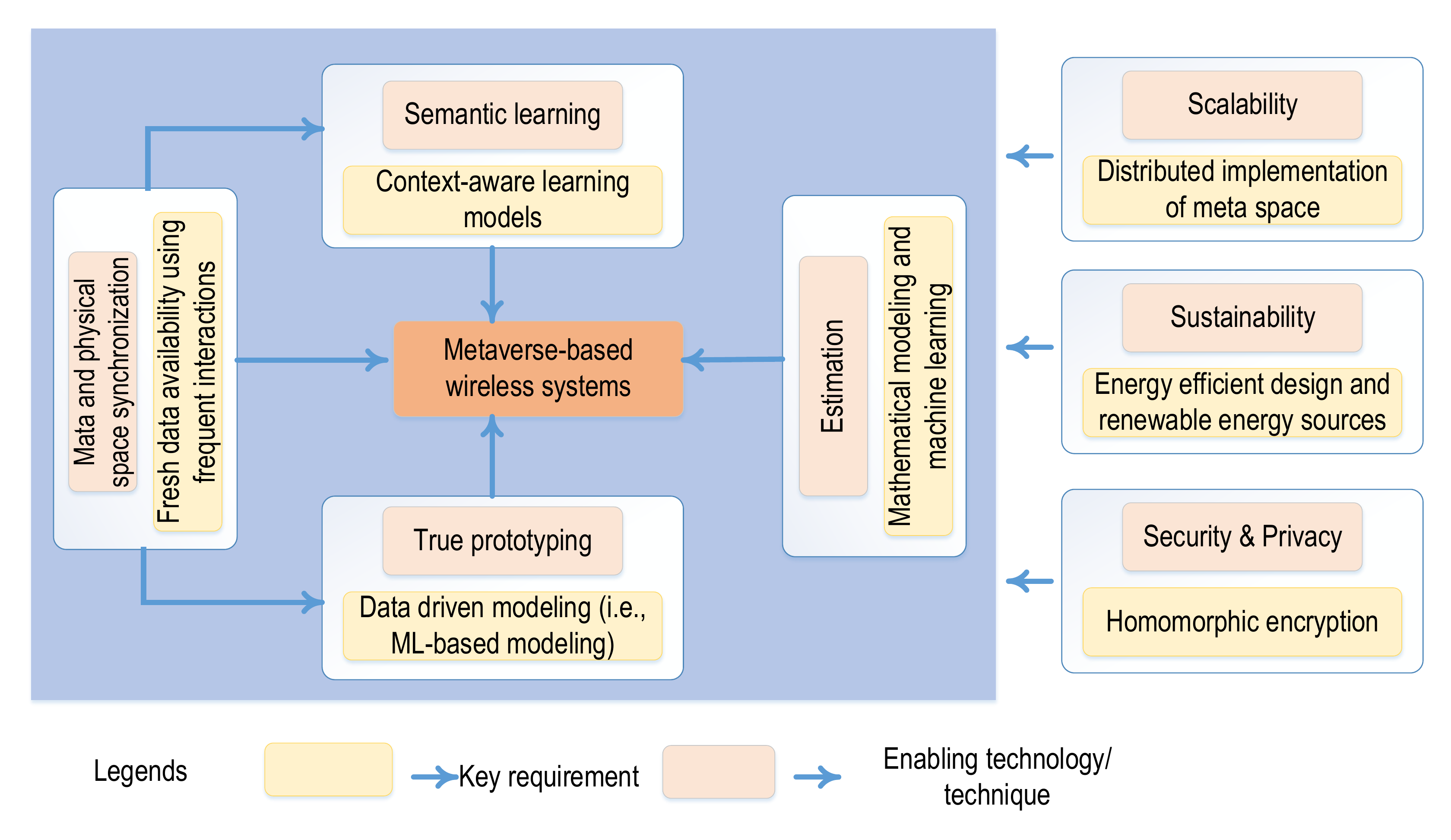}
	\caption{Requirements for metaverse-based wireless systems.}
	\label{fig:requirements}
\end{figure}


\subsubsection{Requirements}
\begin{figure*}[!t]
	\centering
	\captionsetup{justification=centering}
	\includegraphics[width=18cm, height=18cm]{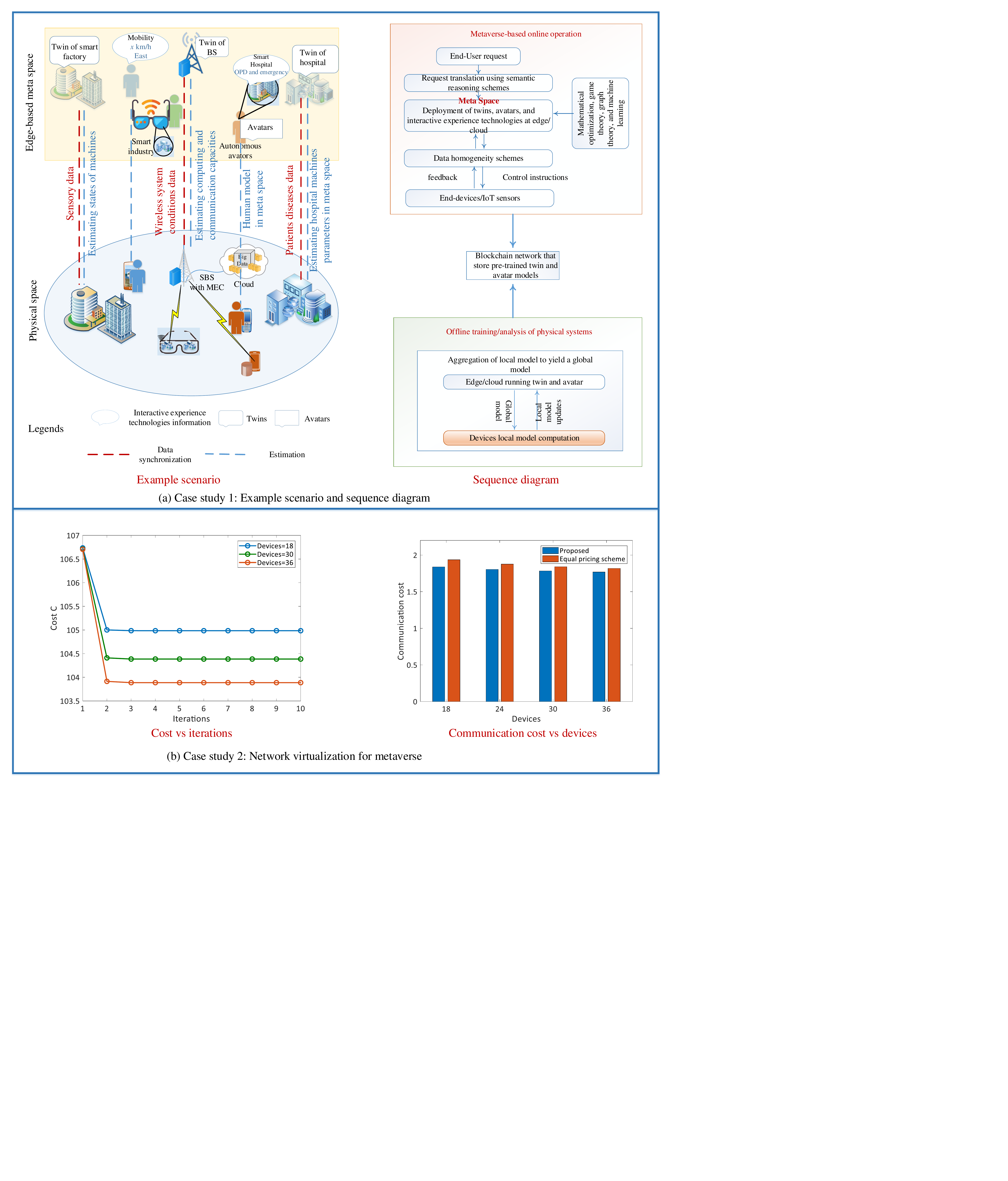}
	\caption{Overview of case studies.}
	\label{fig:example}
\end{figure*}

For a metaverse architecture, general requirements along with relationships with each other are shown in Fig.~\ref{fig:requirements}. These requirements include synchronization between meta space and physical space data  (i.e., fresh data for estimating digital twins and avatars parameters), accurate prototyping of digital twins (e.g., virtual model closest possible to actual physical scenario), scalability (i.e., maintaining the quality of service (QoS) for massive end-devices), sustainability (i.e., energy-efficient operation), semantic reasoning efficiency (i.e., scenario-dependent learning models in wireless communication), security and privacy. In metaverse-based wireless systems (an example scenario is shown in Fig.~\ref{fig:example}, there are two main challenges, namely, synchronization and accurate estimation. Synchronization refers to the data freshness between the meta space and physical space, whereas estimation deals with the accurate representation of digital twins and avatars in meta space. The states of the physical network vary significantly with time. For instance, consider a wireless channel and computing capacity for a certain task (e.g., AR-based industrial management) in the physical system. Other than accurate estimation, data synchronization is necessary for a metaverse-based wireless system. It is necessary to provide fresh data to the metaverse deployed at the network edge/cloud. Such data is needed for performing various tasks such as training machine learning models. For transferring fresh data to edge/cloud-based metaverse, there is a need to allocate dynamically wireless bandwidth. To efficiently carry out data synchronization, the work in \cite{han2021dynamic} proposed a metaverse platform that hosts an infrastructure (e.g., UAVs) for collecting fresh data for multiple virtual service providers (VSPs). A set of UAVs with similar features are used for collecting fresh data for VSPs. To enable such interaction between VSPs and UAVs,  effective incentive mechanisms are required for providing the UAVs with a proportional reward for their sensing tasks. Additionally, there is a need for dynamic resource allocation for performing sensing tasks. On the other hand, as shown in Fig.~\ref{fig:requirements}, semantic learning algorithms, prototyping, meta space and physical space synchronization schemes must be secure, sustainable, and scalable. Scalability enables the efficient operation of a metaverse-based system for a large number of end-users without compromising the QoS. These requirements need to be fulfilled in addition to security for the effective operation of a twin-based system. \par

\subsection{Example Use Cases}
\label{Example Use Case}
\subsubsection{Use Case 1: Metaverse-Empowered Wireless System}
Here, we present a use case of a metaverse-based wireless system as shown in Fig.~\ref{fig:example}. The physical space consists of small cell base stations, edge computing servers, hospitals, industries, mobile users, and a remote cloud. In the meta space, we need twin models of the physical counterparts, such as hospitals, industry, and base stations. Meanwhile, we will also need  to design avatars that effectively model the actual users in the physical space. To give a more concrete example, let us consider an online support service for  a remote monitoring application in an industry $4.0$ scenario which is enabled by metaverse. To manage industrial tasks and faults, the remote live support system will consist of a twin deployed at the network edge, a client with an integrated camera and operating system in the physical interaction space, and the remote expert. The client with a camera captures video and sends it to the twin at the edge server, where computationally expensive tasks occur. Then, the results are sent to the remote expert for further action. The remote expert adds annotations to the video using interactive experience technologies (e.g., XR) and sends them to the client for guidance regarding operation and fault removal. Another example use case can be a \emph{health metaverse} that combines interactive experience technologies with twins and avatars in the meta space to analyze/control healthcare equipment. Such a health metaverse can be used for medical education, training, and surgical procedures.

\subsubsection{Use Case 2: Network Virtualization for Metaverse}
To enable the efficient operation of a metaverse for wireless systems, one can use the concept of network virtualization that enables buying of resources (i.e., computing and wireless) from the network operators and selling them to end-users through metaverse operators. The network operators sell wireless and computing resources and want to maximize their profit, whereas the end-users want to maximize their performance. From the perspective of end-users, one can define a cost function that accounts for wireless resources cost, computing resources cost, communication latency, and computing delay. There is a need to minimize the cost by optimizing the communication resources price, computing resources price, resource allocation, and computing resource management. On the other hand, one must ensure the network operators' minimum acceptable profit using constraints. To carry out the above complex interaction, a metaverse operator will buy resources from the network operators and sell them to the end-users. To do so, We propose an iterative scheme that uses convex optimization for computing resource management, communication resource cost management, and computing resources cost management. For resource allocation, an efficient heuristic algorithm is considered. Fig.~\ref{fig:example} shows the reduction in cost for various number of end-devices. Similarly, the communication resources cost for the proposed and equal pricing scheme is shown in Fig.~\ref{fig:example}, which reveals the performance improvement of our proposal over the equal pricing scheme.     



\section{Conclusion and Future Directions}
We have provided a vision for a metaverse-based wireless system. The metaverse could be a promising technology for  implementing 6G and beyond wireless systems. A set of key enablers for such a  wireless system has been outlined. Additionally, we have proposed a high-level architecture along with an actual wireless system example for a metaverse-based wireless system. several future directions for research are as follows:
\subsection{Meta space and Physical Space Synchronization}
{\em How to achieve efficient and effective synchronization between the meta space and physical space?} In a metaverse-based wireless system, the meta space should be synchronized (i.e., in terms of system states) well with the physical space. To do so, there is a need to propose a joint sampling and communication framework for the metaverse. One should perform effective sampling to effectively collect the system states (e.g., temperature data and vehicular data). Note that sampling uses computing resources that have limitations. Therefore, one must propose a computationally efficient sampling scheme. Next, there is a need for communication and efficient transmission of the sampled data to the meta space. To do so, we should propose quantization (i.e., to minimize the size of the transmitted data) and packet error rate minimization scheme in transmitting the quantized samples to the meta space.

\subsection{Resource Optimization}
{\em How to achieve efficient resource optimization of computing and communication resources for enabling interaction between digital avatars, interactive technologies, and other physical devices?} In a metaverse-based wireless system, computing resources can be at end-devices (i.e. for local training of distributed metaverse models) and edge/cloud server (i.e., for global aggregation of local metaverse models and performing other computing tasks in meta space). Additionally, wireless resources will be required for communication between devices and edge/cloud-based meta space. Therefore, novel resource management schemes will be required for metaverse-based wireless systems.

\subsection{Prototyping}
{\em How to prototype the avatars and digital twins for metaverse?} In a metaverse-based wireless system, modeling digital twins and avatars in a meta space is challenging. For modeling avatars, there is a need to  model physical users' parameters, such as mobility. For mobility,  deep learning can be used to predict the mobility patterns of avatars. Additionally, the estimation of twin models for the wireless system entities will be required. For instance, modeling of the base stations and the propagation channels in meta space can be performed using mathematical modeling, experimental modeling, or data-driven modeling. While data-driven modeling can be more effective than the other techniques, it may incur high training cost. Therefore,  low-complexity machine learning models will be desired for twin modeling.     


\bibliographystyle{IEEEtran}
\bibliography{Database}

\begin{thebibliography}{10}
\providecommand{\url}[1]{#1}
\csname url@samestyle\endcsname
\providecommand{\newblock}{\relax}
\providecommand{\bibinfo}[2]{#2}
\providecommand{\BIBentrySTDinterwordspacing}{\spaceskip=0pt\relax}
\providecommand{\BIBentryALTinterwordstretchfactor}{4}
\providecommand{\BIBentryALTinterwordspacing}{\spaceskip=\fontdimen2\font plus
\BIBentryALTinterwordstretchfactor\fontdimen3\font minus
  \fontdimen4\font\relax}
\providecommand{\BIBforeignlanguage}[2]{{%
\expandafter\ifx\csname l@#1\endcsname\relax
\typeout{** WARNING: IEEEtran.bst: No hyphenation pattern has been}%
\typeout{** loaded for the language `#1'. Using the pattern for}%
\typeout{** the default language instead.}%
\else
\language=\csname l@#1\endcsname
\fi
#2}}
\providecommand{\BIBdecl}{\relax}
\BIBdecl

\bibitem{8869705}
W.~Saad, M.~Bennis, and M.~Chen, ``A vision of 6g wireless systems:
  Applications, trends, technologies, and open research problems,'' \emph{IEEE
  Network}, vol.~34, no.~3, pp. 134--142, 2020.

\bibitem{khan20206g}
L.~U. Khan, I.~Yaqoob, M.~Imran, Z.~Han, and C.~S. Hong, ``6{G} wireless
  systems: A vision, architectural elements, and future directions,''
  \emph{IEEE Access}, vol.~8, pp. 147\,029--147\,044, August 2020.

\bibitem{khan2022digital}
L.~U. Khan, Z.~Han, W.~Saad, E.~Hossain, M.~Guizani, and C.~S. Hong, ``Digital
  twin of wireless systems: Overview, taxonomy, challenges, and
  opportunities,'' \emph{IEEE Communications Surveys \& Tutorials}, 2022.

\bibitem{9711524}
L.~U. Khan, W.~Saad, D.~Niyato, Z.~Han, and C.~S. Hong, ``Digital-twin-enabled
  6g: Vision, architectural trends, and future directions,'' \emph{IEEE
  Communications Magazine}, vol.~60, no.~1, pp. 74--80, 2022.

\bibitem{ning2021survey}
H.~Ning, H.~Wang, Y.~Lin, W.~Wang, S.~Dhelim, F.~Farha, J.~Ding, and
  M.~Daneshmand, ``A survey on the metaverse: The state-of-the-art,
  technologies, applications, and challenges,'' \emph{IEEE Internet of Things
  Journal}, pp. 1--1, 2023.

\bibitem{huynh2022artificial}
T.~Huynh-The, Q.-V. Pham, X.-Q. Pham, T.~T. Nguyen, Z.~Han, and D.-S. Kim,
  ``Artificial intelligence for the metaverse: A survey,'' \emph{Engineering
  Applications of Artificial Intelligence}, vol. 117, p. 105581, 2023.

\bibitem{ynag2022fusing}
Q.~Yang, Y.~Zhao, H.~Huang, Z.~Xiong, J.~Kang, and Z.~Zheng, ``Fusing
  blockchain and ai with metaverse: A survey,'' \emph{IEEE Open Journal of the
  Computer Society}, vol.~3, pp. 122--136, 2022.

\bibitem{chang20226g}
L.~Chang, Z.~Zhang, P.~Li, S.~Xi, W.~Guo, Y.~Shen, Z.~Xiong, J.~Kang,
  D.~Niyato, X.~Qiao \emph{et~al.}, ``6g-enabled edge ai for metaverse:
  Challenges, methods, and future research directions,'' \emph{Journal of
  Communications and Information Networks}, vol.~7, no.~2, pp. 107--121, 2022.

\bibitem{tang2022roadmap}
F.~Tang, X.~Chen, M.~Zhao, and N.~Kato, ``The roadmap of communication and
  networking in 6g for the metaverse,'' \emph{IEEE Wireless Communications},
  2022.

\bibitem{tao2018digital}
F.~Tao, H.~Zhang, A.~Liu, and A.~Y. Nee, ``Digital twin in industry:
  State-of-the-art,'' \emph{IEEE Transactions on Industrial Informatics},
  vol.~15, no.~4, pp. 2405--2415, 2018.

\bibitem{elayan2021digital}
H.~Elayan, M.~Aloqaily, and M.~Guizani, ``Digital twin for intelligent
  context-aware iot healthcare systems,'' \emph{IEEE Internet of Things
  Journal}, vol.~8, no.~23, pp. 16\,749--16\,757, 2021.

\bibitem{khan2021dispersed2}
L.~U. Khan, W.~Saad, Z.~Han, and C.~S. Hong, ``Dispersed federated learning:
  Vision, taxonomy, and future directions,'' \emph{IEEE Wireless
  Communications}, vol.~28, no.~5, pp. 192--198, 2021.

\bibitem{han2019terahertz}
C.~Han, Y.~Wu, Z.~Chen, and X.~Wang, ``Terahertz communications (teracom):
  Challenges and impact on 6g wireless systems,'' \emph{arXiv preprint
  arXiv:1912.06040}, 2019.

\bibitem{salem2018effect}
A.~Salem and M.~M.~A. Azim, ``The effect of rbcs concentration in blood on the
  wireless communication in nano-networks in the thz band,'' \emph{Nano
  communication networks}, vol.~18, pp. 34--43, 2018.

\bibitem{han2021dynamic}
Y.~Han, D.~Niyato, C.~Leung, C.~Miao, and D.~I. Kim, ``A dynamic resource
  allocation framework for synchronizing metaverse with iot service and data,''
  IEEE, pp. 1196--1201, 2022.

\end{thebibliography}

\end{document}